\documentclass[letterpaper, 12pt]{article}

\usepackage{arxiv}
\usepackage{amssymb,amsmath,amsthm}

\usepackage{bm}
\usepackage[colorlinks,citecolor=blue]{hyperref}
\usepackage{subcaption}
\usepackage{multirow}
\usepackage{threeparttable}
\usepackage{booktabs}
\usepackage{rotating}
\usepackage[numbers]{natbib}
\usepackage{dcolumn} 
\usepackage{adjustbox} 

\begin{document}
\title{A generalized double robust Bayesian model averaging approach to causal effect estimation with application to the Study of Osteoporotic Fractures}

\author{Denis Talbot \\ 
     denis.talbot@fmed.ulaval.ca \\
     D\'epartement de m\'edecine sociale et pr\'eventive \\
	Universit\'e Laval \\
	\And
  Claudia Beaudoin\\
 claudia.beaudoin@crchudequebec.ulaval.ca \\
D\'epartement de m\'edecine sociale et pr\'eventive \\
	Universit\'e Laval \\
}

\maketitle

\begin{abstract}
Analysts often use data-driven approaches to supplement their substantive knowledge when selecting covariates for causal effect estimation. Multiple variable selection procedures tailored for causal effect estimation have been devised in recent years, but additional developments are still required to adequately address the needs of data analysts. In this paper, we propose a Generalized Bayesian Causal Effect Estimation (GBCEE) algorithm to perform variable selection and produce double robust estimates of causal effects for binary or continuous exposures and outcomes. GBCEE employs a prior distribution that targets the selection of true confounders and predictors of the outcome for the unbiased estimation of causal effects with reduced standard errors. Double robust estimators provide some robustness against model misspecification, whereas the Bayesian machinery allows GBCEE to directly produce inferences for its estimate. GBCEE was compared to multiple alternatives in various simulation scenarios and was observed to perform similarly or to outperform double robust alternatives. Its ability to directly produce inferences is also an important advantage from a computational perspective. The method is finally illustrated for the estimation of the effect of meeting physical activity recommendations on the risk of hip or upper-leg fractures among elderly women in the Study of Osteoporotic Fractures. The 95\% confidence interval produced by GBCEE is 61\% shorter than that of a double robust estimator adjusting for all potential confounders in this illustration. 
\end{abstract}

\quad

\begin{center}
\textbf{Keywords} : Causal inference; Confounding; Double robustness; Model averaging; Model selection;
\end{center}

\clearpage
\section{Introduction}
\label{s:intro}

Estimating causal effects using observational data requires important subject-matter knowledge. One notably needs to identify the confounding covariates that bias the association between the exposure and the outcome. Unfortunately, prior knowledge is often insufficient to undertake this task. For example, reviews of variable selection methods used in epidemiological journals indicate that data-driven approaches are frequently employed to select confounders \cite{Walter, Talbot2019}. 

Recent research has shown that many classical model selection approaches, including stepwise regression, Bayesian model averaging, lasso and adaptive lasso, can have poor performances in a causal inference framework (for example, \cite{Ertefaie, Shortreed, Talbot2015, Wang2012}). There are two important explanations of this phenomenon. First, many model selection approaches do not account for the model selection steps and therefore produce confidence intervals that include the true effect less often then they should. Second, classical methods often fail to identify all important confounders. This is because they generally focus on modeling the outcome as a function of the exposure and potential confounders. Confounders that are weakly associated with the outcome but strongly with the exposure may thus fail to be identified. 

Multiple model selection methods for causal inference have been introduced in recent years, including the collaborative targeted maximum likelihood (C-TMLE; \cite{vanderLaan2010}), Bayesian Adjustment for Confounding (BAC; \cite{Wang2012, Wang2015}), Confounder selection via penalized credible regions (BP; \cite{Wilson}), Inference on treatment effects after selection among high-dimensional controls (HDM; \cite{Belloni}), Bayesian Causal Effect Estimation algorithm (BCEE; \cite{Talbot2015}), Model averaged double robust estimation (MADR; \cite{Cefalu}), Outcome-adaptive Lasso (OAL; \cite{Shortreed}), and the Group lasso and double robust estimation of causal effects (GLiDeR; \cite{Koch}). Despite these important developments, there is still a need for new methods or for the extension of existing methods. For instance, not all methods directly produce standard errors or confidence intervals for their causal effect estimator. Instead, they rely on bootstrap procedures for making inferences, which can be computationally intensive. Moreover, many methods were specifically elaborated for estimating the effect of a binary exposure on a continuous outcome. Only a few address additional situations. Finally, most methods either model the outcome as a function of the exposure and confounders, or the exposure as a function of confounders, and require that the postulated model is correct. In contrast, double robust methods combine both models and yield unbiased estimators if either model, but not necessarily both, is correct. Table \ref{table1} presents a summary of the characteristics of the causal model selection methods mentioned earlier. 

\begin{table} [hb!]
\caption{Characteristics of causal model selection methods}
\label{table1}
\begin{small}
\begin{center}
\begin{tabular}{cccccccc}
\hline
Method & R       & Binary   & Continuous  & Binary  & Continuous    & Modeling & Inferences \\
       & package & exposure & exposure    & outcome & outcome             &          & \\
 \hline
C-TMLE & \texttt{ctmle} & \checkmark &   & \checkmark & \checkmark & Double robust & Asymptotic \\
BAC & \texttt{bacr}, \texttt{BACprior} & \checkmark & \checkmark & \checkmark & \checkmark & Outcome & Bayesian \\
BP & \texttt{BayesPen}$^1$ & \checkmark & \checkmark & \checkmark & \checkmark & Outcome & Bayesian \\
HDM & \texttt{hdm} & \checkmark &  &  & \checkmark & Outcome & Asymptotic \\
BCEE & \texttt{BCEE} & \checkmark & \checkmark &  & \checkmark & Outcome & Bayesian \\
MADR & \texttt{madr} & \checkmark &   & $\pm^2$ & \checkmark & Double robust & Bootstrap \\
OAL & None$^3$ & \checkmark &   & \checkmark & \checkmark & Exposure & Bootstrap \\
GLiDeR & None$^4$ & \checkmark &   & $\pm^2$ & \checkmark & Double robust & Bootstrap \\
\hline
\end{tabular}
\end{center}
$^1$ The \texttt{BayesPen} package was removed from the CRAN repository in December 2014 because the maintainer email address bounced.\\
$^2$ The proposed method would directly adapt to this situation, but the software does not accommodate it.\\
$^3$ R code is provided in \cite{Shortreed}, but the code depends on the \texttt{lqa} R package that has been removed from the CRAN repository due to uncorrected error checks. The \texttt{lqa} package is still available on the CRAN archives. \\
$^4$ R code is provided in \cite{Koch}.
\end{small}
\end{table}

In this paper, we extend the BCEE algorithm to select confounders and produce double robust causal effect estimation for binary or continuous exposure and outcome. We call this extension the Generalized Bayesian Causal Effect Estimation algorithm (GBCEE). BCEE's framework has various desirable features. First, its model selection algorithm is theoretically motivated using the graphical framework to causal inference (for example, \cite{Pearl2009}). This algorithm favors models that include true confounders in addition to outcome risk factors, but that exclude instruments (covariates associated with the exposure, but not the outcome). Simulation studies indicate that such models unbiasedly estimate causal effects with improved efficiency as compared to models including all potential confounders or to models including only true confounders \cite{Brookhart, deLuna, Shortreed}. Moreover, BCEE takes advantage of the Bayesian framework to directly produce inferences that account for the model selection step, thus avoiding reliance on the bootstrap. All extensions presented in this article will be implemented in the R package \texttt{BCEE}. 

The remainder of the paper is structured as follows. In Section \ref{notation}, we introduce some notation and motivate the confounder selection problem. The GBCEE algorithm is presented in Section \ref{algo}. Section \ref{simul} presents a simulation study to investigate and compare the finite sample properties of GBCEE. An illustration of GBCEE's application for estimating the effect of physical activity on the risk of fractures using data from the Study of Osteoporotic Fractures is presented in Section \ref{application}. We conclude in Section \ref{s:discuss} with a discussion of the results and perspectives for future research.

\section{Notation and motivation}
\label{notation}

Let $Y$ be the outcome of interest and $X$ be the exposure under study. Let $Y^x$ be the counterfactual outcome that would have been observed if, possibly contrary to fact, $X = x$. We denote by $\Delta$ the causal exposure effect of interest. We restrict our attention to causal effects that can be expressed as contrasts between two counterfactual expectations of the outcome, $\mathbb{E}[Y^x]$ vs $\mathbb{E}[Y^{x'}]$, for $x \neq x'$. For example, if both exposure and outcome are binary, $\Delta$ can be the causal risk difference $\mathbb{E}[Y^1] - \mathbb{E}[Y^0]$. 

For estimating the causal effect, a sample of independent and identically distributed observations $\{\bm{U}_i, X_i, Y_i\}$, $i = 1, ..., n$, is drawn from a given population, where $\bm{U}_i = \{U_{i,1}, ..., U_{i,M}\}$ is a set of potentially confounding covariates. We assume conditional exchangeability $Y^x \coprod X | \bm{U}$ and positivity $f_X(x|\bm{U} = \bm{u}) > 0$ for all $\bm{u}$ and all $x$ where $f_X(x) > 0$. The conditional exchangeability assumption is often interpreted as a ``no unmeasured confounders'' assumption and informally entails there is no factor that simultaneously affects $X$ and $Y$ after conditioning on $\bm{U}$. The positivity assumption means that all possible values of $X$ have a positive probability of occurring within each stratum of $\bm{u}$. Under these assumptions, counterfactual expectations $\mathbb{E}[Y^x]$ are non parametrically identified from the observed data as $\mathbb{E}_{\bm{U}}\left\{\mathbb{E}\left[Y|X=x, \bm{U}\right] \right\}$. Confounder selection entails finding a subset of $\bm{U}$, $\bm{Z} \subseteq \bm{U}$, such that the conditional exchangeability assumption also holds, $Y^x \coprod X | \bm{Z}$. 

Confounder selection among the set $\bm{U}$ can be desirable for multiple reasons. When $M\approx n$ or $M > n$, it might not be feasible to estimate ${\mathbb{E}[Y|X=x, \bm{U}]}$ because of data sparsity. Moreover, $\bm{U}$ is susceptible to include instruments or conditional instruments. Adjusting for such variables is known to increase the variance of estimators and is susceptible to amplify biases (for example, see \cite{Pearl2011} and references therein). Finally, identifying which covariates are true confounders may be of substantive interest.

\section{The GBCEE algorithm}
\label{algo}

We now describe the GBCEE algorithm. We first provide an overview of the algorithm in the next subsection. The construction of the prior distribution is presented in Section~\ref{prior}. The double robust estimators used within the algorithm are presented in Section~\ref{estimators}. The implementation of the algorithm is described in Section~\ref{implementation}.

\subsection{Overview}
\label{overview}

GBCEE averages double robust estimates of the causal effect $\Delta$ over potential confounder sets. These sets are the $2^M$ possible subsets of $\bm{U}$. Let $\bm{\alpha^Y} = (\alpha^Y_1, ..., \alpha^Y_M)$ be an $M$-dimensional vector for the inclusion of the covariates $\bm{U}$ in the potential confounder set, where component ${\alpha^Y_m}$ equals $1$ if $U_m$ is included in the set and $\alpha^Y_m$ equals $0$ otherwise. Let $\hat{\Delta}^{\bm{\alpha^Y}}$ be the double robust estimate of $\Delta$ produced by considering the set defined by $\bm{\alpha^Y}$ and $\mathcal{A}$ be the set of all possible confounder sets. As will be seen in Section \ref{estimators}, each $\hat{\Delta}^{\bm{\alpha^Y}}$ is constructed using the output of an exposure model and an outcome model, both of which include the potential confounders indicated by $\bm{\alpha^Y}$. Using this notation, GBCEE estimates $\Delta$ as
\begin{equation*}
\hat{\Delta} = \mathbb{E}[\Delta|Y] = \sum_{\bm{\alpha^Y} \in \mathcal{A}} \hat{\Delta}^{\bm{\alpha^Y}}P(\bm{\alpha^Y}|Y).
\end{equation*}   

The weight attributed to each estimate $\hat{\Delta}^{\bm{\alpha^Y}}$ is the posterior probability of a model for the outcome, conditional on the exposure and the covariates included in $\bm{\alpha^Y}$, ${P(\bm{\alpha^Y}|Y) \propto P(Y|\bm{\alpha^Y})P(\bm{\alpha^Y})}$. We propose using a generalized linear model as the model for the outcome:
\begin{gather}
g\left[\mathbb{E}(Y|\bm{\alpha^Y}, X, \bm{U})\right] = \delta_0^{\bm{\alpha^Y}} + \beta^{\bm{\alpha^Y}} X + \sum_{m = 1}^{M}{\alpha_m^Y \delta_m^{\bm{\alpha^Y}} U_{m}}, \label{glm_outcome_model}
\end{gather}

\noindent where $g$ is a known link function. This outcome model serves as component in constructing $\hat{\Delta}^{\bm{\alpha^Y}}$, in addition to being used for computing $P(Y|\bm{\alpha^Y})$. 

So far, GBCEE is thus identical to a classical Bayesian model averaging of $\hat{\Delta}^{\bm{\alpha^Y}}$. The important differentiating feature of GBCEE is that it uses an informative prior distribution, $P(\bm{\alpha^Y})$, tailored to put the bulk of the posterior weight on potential confounder sets that meet the conditional exchangeability assumption. The prior distribution is further constructed to favor sets that exclude instruments or conditional instruments. As such, the prior distribution targets sets $\bm{\alpha^Y}$ that allow unbiased estimation of $\Delta$, with improved efficiency as compared to the full set $\bm{U}$.

\subsection{Prior distribution construction}
\label{prior}

The first step for constructing the prior distribution $P(\bm{\alpha^Y})$ aims to identify the determinants of the exposure. This step helps identifying instruments, conditional instruments as well as confounders in later steps. To do so, we consider generalized linear models for the exposure according to potential confounders. Let $\bm{\alpha^X} \in \mathcal{A}$ be defined similarly to $\bm{\alpha^Y}$, that is, $\bm{\alpha^X}$ is an $M$-dimensional vector for the inclusion of the covariates $\bm{U}$ in the exposure model. The exposure models are 
\begin{equation}
g\left[\mathbb{E}(X|\bm{\alpha^X}, \bm{U})\right] = \delta_0^{\bm{\alpha^X}} + \sum_{m = 1}^{M}{\alpha_m^X \delta_m^{\bm{\alpha^X}} U_{m}}, \label{exposure_model}
\end{equation}
\noindent where $g$ is a known link function. The posterior distribution of the exposure model $P(\bm{\alpha^X}|X) \propto P(X|\bm{\alpha^X})P(\bm{\alpha^X})$ is then computed. Typically, a uniform prior distribution over $\mathcal{A}$ would be used for $P(\bm{\alpha^X})$. The set $\bm{\alpha^X}$ that includes all true predictors of the exposure is ensured to have a posterior probability of 1 asymptotically if the exposure model is correctly specified \cite{Haughton, Wasserman}. As described in Section \ref{estimators}, the exposure models $(\ref{exposure_model}$) are also used for constructing the double robust estimates $\Delta^{\bm{\alpha^Y}}$. 

Using the information from the first step, GBCEE's prior distribution is defined as ${P(\bm{\alpha^Y}) = \sum_{\bm{\alpha^X} \in \mathcal{A}} P\left(\bm{\alpha^Y}| \bm{\alpha^X}\right) P(\bm{\alpha^X}|X)}$, where $P\left(\bm{\alpha^Y}| \bm{\alpha^X}\right) \propto \prod_{m = 1}^M P_{\bm{\alpha^Y}|\bm{\alpha^X}}\left(\alpha_m^Y|\alpha_m^X\right)$, and
\begin{gather*}
P_{\bm{\alpha^Y}|\bm{\alpha^X}}\left(\alpha_m^Y = 1  | \alpha_m^X = 0\right) = P_{\bm{\alpha^Y}|\bm{\alpha^X}}\left(\alpha_m^Y = 0 | \alpha_m^X = 0\right) = \frac{1}{2}, \\ 
P_{\bm{\alpha^Y}|\bm{\alpha^X}}\left(\alpha_m^Y = 1  | \alpha_m^X = 1\right)  = \frac{\omega^{\bm{\alpha^Y}}_m}{\omega^{\bm{\alpha^Y}}_m + 1}, \\ 
P_{\bm{\alpha^Y}|\bm{\alpha^X}}\left(\alpha_m^Y = 0 | \alpha_m^X = 1\right) = \frac{1}{\omega^{\bm{\alpha^Y}}_m + 1}.
\end{gather*}   

\noindent Intuitively, if $U_m$ was not found to be associated with the exposure in the first step of GBCEE ($\alpha_m^X = 0$), there is no reason to force either the inclusion or the exclusion of $U_m$ from the outcome model, hence $P_{\bm{\alpha^Y}|\bm{\alpha^X}}\left(\alpha_m^Y = 1  | \alpha_m^X = 0\right) = P_{\bm{\alpha^Y}|\bm{\alpha^X}}\left(\alpha_m^Y = 0 | \alpha_m^X = 0\right) = 1/2$. However, if $U_m$ was found to be associated with the exposure ($\alpha_m^X = 1$), it is desirable (i) to force its inclusion in the outcome model if $U_m$ is also associated with the outcome ($U_m$ is a confounder), and (ii) force its exclusion from the outcome model if $U_m$ is not associated with the outcome ($U_m$ is an instrument or a conditional instrument). This is accomplished through the parameter $\omega^{\bm{\alpha^Y}}_m$, which is proportional to the strength of the association between $U_m$ and $Y$ conditional on the other variables in the outcome model $\bm{\alpha^Y}$. Some additional notation must be introduced to define $\omega^{\bm{\alpha^Y}}_m$ formally.  

We denote by $\tilde{\delta}_m^{\bm{\alpha^Y}}$ the regression coefficient for $U_m$ in the outcome model defined by $\bm{\alpha^Y}$ if $U_m$ is included in this model (that is, if $\alpha^Y_m$ = 1). Otherwise, $\tilde{\delta}_m^{\bm{\alpha^Y}}$ is the regression coefficient for $U_m$ in the outcome model that includes the same covariates as $\bm{\alpha^Y}$, but additionally include $U_m$. Using this notation, 
\begin{equation}
\omega^{\bm{\alpha^Y}}_m = \omega \times \left(\tilde{\delta}_m^{\bm{\alpha^Y}}\frac{\sigma_{U_m}}{\sigma_Y}\right)^2,
\label{omega}
\end{equation}

\noindent where ${0 \leq \omega \leq \infty}$ is a user-defined hyperparameter, $\sigma_{U_m}$ and $\sigma_Y$ are the standard deviations of $U_m$ and $Y$, respectively. Note that the term $\sigma_{U_m}/\sigma_Y$ makes $\omega^{\bm{\alpha^Y}}_m$ insensitive to the choice of scale for $Y$ and $U_m$. However, when $Y$ is binary (0/1), the regression coefficients do not need to be scaled for $\sigma_Y$; we thus set $\sigma_Y = 1$ in Equation (\ref{omega}) in that case. Since $\tilde{\delta}_m^{\bm{\alpha^Y}}$, $\sigma_{U_m}$ and $\sigma_Y$ are unknown, they are replaced by their maximum likelihood estimates. As will be seen in Section \ref{implementation}, the uncertainty associated with this estimation is accounted for in the implementation of GBCEE. It is recommended to choose $\omega = c\times n^b$, with $0<b<1$ and $c$ is a constant that does not depend on sample size. This ensures that the prior distribution behaves as desired asymptotically, that is, that confounders are included and instruments and conditional instruments are excluded \cite{Talbot2015}. More details concerning the prior distribution can be found in \cite{Talbot2015}. 

\subsection{Double robust estimators}
\label{estimators}

Targeted maximum likelihood estimation (TMLE) is a general framework for constructing double robust estimators introduced by \cite{vanderLaan2006}, but based on earlier work \cite{Scharfstein}. TMLE is consistent if either the outcome or the exposure model is correctly specified, and is semiparametric efficient when both are correctly specified. It has also been suggested that TMLE is robust to near violations of the positivity assumption \cite{Porter, vanderLaan2011}. For these reasons, we propose using TMLE in GBCEE.

The estimators we propose using depend on the type of the outcome and of the exposure. For constructing these estimators, we use the generalized linear model (\ref{glm_outcome_model}) for the outcome with an identity link when $Y$ is continuous and with a logit link when $Y$ is binary. For the exposure model, we use the model (\ref{exposure_model}) with an identity link when $X$ is continuous and with a logit link when $X$ is binary. 

Briefly, in our case, TMLE first entails proposing an initial estimate for the causal contrast of interest based on the outcome model. The initial estimate is then fluctuated based on the output from the exposure model. The intuition is that the exposure model should contain no additional information for predicting the outcome if the initial outcome model is correctly specified. A residual association thus suggests that residual confounding is present. TMLE uses this residual association to fluctuate the initial estimate in a clever way, that is, such that the estimating equations of the efficient influence function of the causal contrast is solved. The efficient influence function is an analogue to the Cram\'er-Rao variance lower bound for semiparametric estimators. A more technical presentation of TMLE can be found in \cite{vanderLaan2011}. 

We now briefly expose the contrasts of interest and the TMLE estimators we propose using. More details are provided in Web Appendix~A. When both $Y$ and $X$ are continuous, the causal effect of interest is $\Delta = \mathbb{E}[Y^{x+1}] - \mathbb{E}[Y^{x}]$, assuming a linear effect of $X$ on $Y$. The estimator we use is the TMLE proposed in Chapter 22 of \cite{vanderLaan2011}. When $Y$ is continuous and $X$ is binary the targeted causal effect is $\Delta = \mathbb{E}[Y^1] - \mathbb{E}[Y^0]$ and the estimator is the TMLE proposed in Chapter 4 of \cite{vanderLaan2011}. We use the same algorithm as \cite{Siddique} for estimating $\mathbb{E}[Y^x]$, for $x = 0, 1$. When both $Y$ and $X$ are binary, we consider either $\Delta = \mathbb{E}[Y^1] - \mathbb{E}[Y^0]$ or $\Delta = \mathbb{E}[Y^1]/\mathbb{E}[Y^0]$. The same method as in the previous case is employed for estimating $\mathbb{E}[Y^x]$. Finally, when $Y$ is binary and $X$ is continuous, we consider $\Delta = \mathbb{E}[Y^x] - \mathbb{E}[Y^{x'}]$ or $\Delta = \mathbb{E}[Y^{x}]/\mathbb{E}[Y^{x'}]$, with $x \neq x'$. We use the TMLE proposed in Chapter 9 of \cite{vanderLaan2011}, initially introduced by \cite{Rosenblum}. 

\subsection{Implementation}
\label{implementation}

The proposed algorithm for sampling in the posterior distribution of $\Delta$ is similar to the one proposed by \cite{Talbot2015}, but has been revised to improve computational efficiency.

\textit{Step 1.} Determine the posterior distribution of the exposure model $P(\bm{\alpha^X}|X) \propto P(X|\bm{\alpha^X})$. Efficient algorithms for performing this task are readily available in usual software, for example in the package BMA in R \cite{Raftery2018}.

\textit{Step 2.} Use a Markov chain Monte Carlo model composition algorithm to explore the outcome model space $\bm{\alpha^Y}$ \cite{Madigan}. This algorithm starts with an initial outcome model. At each step of the algorithm, a candidate model is randomly chosen by adding or removing one covariate $U_m$ from the outcome model. The decision to move to the candidate model  ($\bm{\alpha^Y_1}$) or to remain in the current model ($\bm{\alpha^Y_2}$) is based on the Metropolis-Hasting ratio of the posterior probability of the outcome models 
\begin{align*}
\frac{P(\bm{\alpha^Y_1}|Y)}{P(\bm{\alpha^Y_2}|Y)}&= \frac{P(Y|\bm{\alpha^Y_1})}{P(Y|\bm{\alpha^Y_2})} \times \frac{\sum_{\bm{\alpha^X}} P\left(\bm{\alpha^Y_1}| \bm{\alpha^X}\right) P(\bm{\alpha^X}|X)}
{\sum_{\bm{\alpha^X}} P\left(\bm{\alpha^Y_2}| \bm{\alpha^X}\right)P(\bm{\alpha^X}|X)} \nonumber \\ 
& = \frac{P(Y|\bm{\alpha^Y_1})}{P(Y|\bm{\alpha^Y_2})} \times \frac{\sum_{\bm{\alpha^X}}\prod_{m = 1}^M P_{\bm{\alpha^Y_1}|\bm{\alpha^X}}\left(\alpha_m^Y|\alpha_m^X\right)P(\bm{\alpha^X}|X)}
{\sum_{\bm{\alpha^X}} \prod_{m = 1}^M P_{\bm{\alpha^Y_2}|\bm{\alpha^X}}\left(\alpha_m^Y|\alpha_m^X\right)P(\bm{\alpha^X}|X)}.
\end{align*}

Similarly to \cite{Talbot2015}, we propose the following approximation 
\begin{align*}
 \frac{\sum_{\bm{\alpha^X}}\prod_{m = 1}^M P_{\bm{\alpha^Y_1}|\bm{\alpha^X}}\left(\alpha_m^Y|\alpha_m^X\right)P(\bm{\alpha^X}|X)}
{\sum_{\bm{\alpha^X}} \prod_{m = 1}^M P_{\bm{\alpha^Y_2}|\bm{\alpha^X}}\left(\alpha_m^Y|\alpha_m^X\right)P(\bm{\alpha^X}|X)}  \approx \frac{\sum_{\bm{\alpha^X}} P_{\bm{\alpha^Y_1}|\bm{\alpha^X}}\left(\alpha_m^Y|\alpha_m^X\right)P(\bm{\alpha^X}|X)}
{\sum_{\bm{\alpha^X}} P_{\bm{\alpha^Y_2}|\bm{\alpha^X}}\left(\alpha_m^Y|\alpha_m^X\right)P(\bm{\alpha^X}|X)}.
\end{align*}
The intuition of this approximation is that the two outcome models $\bm{\alpha^Y_1}$ and $\bm{\alpha^Y_2}$ only differ in their inclusion of covariate $U_m$. As such, it is expected that the term related to $U_m$ in $\prod_{m = 1}^M P_{\bm{\alpha^Y}|\bm{\alpha^X}}\left(\alpha_m^Y|\alpha_m^X\right)$ is the most influential on the value of the ratio.

Since $P_{\bm{\alpha^Y}|\bm{\alpha^X}}\left(\alpha_m^Y|\alpha_m^X\right)$ depends on $\tilde{\delta}_m^{\bm{\alpha^Y}}$ which is unknown, we proposed replacing $\tilde{\delta}_m^{\bm{\alpha^Y}}$ by its maximum likelihood estimate. To account for the uncertainty in this estimate, we can write $P_{\bm{\alpha^Y}|\bm{\alpha^X}}(\alpha^Y_m|\alpha^X_m) = \int_{\tilde{\delta}_m^{\bm{\alpha^Y}}} P_{\bm{\alpha^Y}|\bm{\alpha^X}}(\alpha^Y_m|\alpha^X_m, \tilde{\delta}_m^{\bm{\alpha^Y}}) f(\tilde{\delta}_m^{\bm{\alpha^Y}}) d\tilde{\delta}_m^{\bm{\alpha^Y}}$  where $f(\tilde{\delta}_m^{\bm{\alpha^Y}})$ is approximated by its limit distribution, the maximum likelihood estimator distribution $N\left(\hat{\tilde{\delta}}_m^{\bm{\alpha^Y}}, \widehat{SE}(\hat{\tilde{\delta}}_m^{\bm{\alpha^Y}})\right)$ \cite{Walker, Dawid}. 

\textit{Step 3.} For each $\bm{\alpha^Y}$ explored, compute $\hat{\Delta}^{\bm{\alpha^Y}}$, $\widehat{Var}(\hat{\Delta}^{\bm{\alpha^Y}})$ and $P(\bm{\bm{\alpha^Y}}|Y)$. 

The causal contrasts $\hat{\Delta}^{\bm{\alpha^Y}}$ are estimated using the double robust estimators presented in the previous section. An asymptotic estimator for the variance of these estimators is the sample variance of the efficient influence function divided by $n$ \cite{vanderLaan2011}. The efficient influence functions of the estimators can be found in their respective references given previously (see Section \ref{estimators}). While this variance estimator is computationally attractive, it is consistent only if both the exposure and the outcome models are correctly specified, and is conservative if only the exposure model is correctly specified. An alternative solution is using the nonparametric bootstrap. This may be seen as computationally undesirable, but it is important to remark that the bootstrap is performed within, and not over, the GBCEE algorithm, thus reducing the computational cost.

The posterior probability of each outcome model $P(\bm{\alpha^Y}|Y)$ can be approximated using the Metropolis-Hasting ratios of the posterior probability of the outcome models calculated in Step 2. Remark that it is possible to obtain ratios comparing the posterior probability of each explored model to a single alternative model, for example $\bm{\alpha^Y_1}$, by multiplying together ratios comparing different models. For example, $\frac{P(\bm{\alpha^Y_3}|Y)}{P(\bm{\alpha^Y_1}|Y)} = \frac{P(\bm{\alpha^Y_3}|Y)}{P(\bm{\alpha^Y_2}|Y)} \times \frac{P(\bm{\alpha^Y_2}|Y)}{P(\bm{\alpha^Y_1}|Y)}$. The posterior probability of each model is then obtained by dividing each of these ratios by their sum. 

\textit{Step 4.} Compute the posterior expectation $\mathbb{E}(\Delta|Y) = \sum_{\bm{\alpha^Y}} \hat{\Delta}^{\bm{\alpha^Y}} P(\bm{\alpha^Y}|Y)$ and the posterior variance $Var(\Delta|Y)= \sum_{\bm{\alpha^Y}} \left[ \widehat{Var}(\hat{\Delta}^{\bm{\alpha^Y}}) + (\hat{\Delta}^{\bm{\alpha^Y}})^2 \right]P(\bm{\alpha^Y}|Y) - \mathbb{E}(\Delta|Y)^2.$ \cite{Raftery}

\section{Simulation study} \label{simul}

\subsection{Scenarios}
We consider 5 scenarios where the exposure is binary and the outcome is continuous. Two of these scenarios are then adapted to the case where the outcome is instead binary. We focus on the binary exposure -- continuous outcome case since it is the case for which the most comparators are available. 

Scenario 1 is inspired from \cite{Talbot2015}. The data-generating equations are $U_{6},..., U_{40} \stackrel{iid}{\sim}N(0, 1)$, $U_{1}, ..., U_{5} \stackrel{iid}{\sim}N(U_{11} + ... + U_{15}, 1)$, $X \sim Bernoulli(p = expit(U_{11} + ... + U_{30}))$ and $Y \sim N(X + 0.1U_{1} + ... + 0.1U_{10}, 1)$, where $expit(x) = [1 + exp(-x)]^{-1}$. 

Scenario 2 is taken from \cite{Shortreed}. $U_1, ..., U_{20}$ were generated as multivariate normal variables with mean 0, unit variance and 0.5 correlation, $X \sim Bernoulli(p = expit(U_1 + U_2 + U_5 + U_6))$ and $Y \sim N(2X + 0.6 U_1 + 0.6 U_2 + 0.6 U_3 + 0.6 U_4, 1)$. 

Scenario 3 is taken from \cite{Ertefaie}. The data were generated as $U_1, ..., U_{100} \stackrel{iid}{\sim} N(1, \sigma^2 = 4)$, $X \sim Bernoulli(p = expit(0.5 U_1 - U_2 + 0.3 U_5 - 0.3 U_6 + 0.3 U_7 - 0.3 U_8))$, $Y \sim N(X + 2U_1 + 0.2U_2 + 5U_3 + 5U_4, \sigma^2 = 4)$. 

Scenarios 4 and 5 are taken from \cite{Koch}. In both scenarios, $U_1, ..., U_5$ were generated as multivariate normal variables with mean 1, unit variance and 0.6 correlations. In Scenario~4, $X \sim Bernoulli(p = expit(0.5 U_1 + 0.5 U_2 + 0.1 U_3))$ and $Y \sim N(X + U_3 + U_4 + U_5 + \sum_{i = 1}^5 \sum_{j = 1}^5 0.5 U_i U_j, 1)$. In Scenario~5, $X \sim Bernoulli(p = expit(-5 + U_3 + U_4 + U_5 + \sum_{i = 1}^5 \sum_{j = 1}^5 0.5 U_i U_j))$, and $Y\sim N(X + 0.5 U_1 + 0.5 U_2 + 0.1 U_3, 1)$. 

Scenarios 2 and 4 are adapted to the binary outcome (Scenarios 2B and 4B). Only the equation used for generating $Y$ were changed; they were, respectively, $Y \sim Bernoulli(p = expit(2X + 0.6 U_1 + 0.6 U_2 + 0.6 U_3 + 0.6 U_4))$ and $Y \sim Bernoulli(p = expit(-5 + X + U_3 + U_4 + U_5 + \sum_{i = 1}^5 \sum_{j = 1}^5 0.5 U_i U_j))$. 

In Scenarios~1-5, both a sample size of $n = 200$ and $n = 1000$ were considered. In Scenarios~2B and 4B, only a sample size of $n = 1000$ was considered because of frequent convergence problems with $n = 200$, due to too few events relative to the number of variables.  

\subsection{Analysis}

For each scenario, 1000 datasets were generated. The estimand of interest was the average treatment effect $\Delta = \mathbb{E}[Y^1] - \mathbb{E}[Y^0]$. In Scenarios~1-5, $\Delta$ was estimated with GBCEE setting $\omega = 500 \sqrt{n}$, OAL, C-TMLE, GLiDeR, BAC, MADR, BP, and HDM. For Scenarios 2B and 4B, only GBCEE, OAL, C-TMLE and BAC were considered. The other methods, or the software available for these methods, do not allow estimating the average treatment effect for binary outcomes. 

As benchmarks, in all scenarios, we have additionally estimated $\Delta$ using the parametric g-formula $\hat{\Delta}_g = \frac{1}{n}\sum_{i = 1}^n( \hat{\mathbb{E}}[Y|X = 1, \bm{U}_i] - \hat{\mathbb{E}}[Y|X = 0, \bm{U}_i])$ and with a double robust augmented inverse probability weighting (AIPW) estimator:
\begin{align*}
\hat{\Delta}_{AIPW} & =\frac{1}{n}\sum_{i = 1}^n \left\{\left[\frac{X_i}{\hat{P}(X = 1|\bm{U}_i)} + \frac{1 - X_i}{1 - \hat{P}(X = 1|\bm{U}_i)}\right] \right. \times \\
& \left. \left[Y_i - \hat{\mathbb{E}}(Y|X_i,\bm{U}_i)\right] + \hat{\mathbb{E}}(Y|X = 1,\bm{U}_i) - \hat{\mathbb{E}}(Y|X = 0,\bm{U}_i)\right\},
\end{align*}
where $\hat{\mathbb{E}}[Y|X, \bm{U}]$ was obtained from a linear regression model in Scenarios~1-5 and a logistic regression model in Scenarios~2B and 4B, and $\hat{P}(X = 1|\bm{U}_i)$ was produced by a logistic regression model. These regression models were either adjusted for all potential confounders (full-g and full-AIPW, respectively) or only for the pure outcome predictors and true confounders (target-g and target-AIPW, respectively). The target adjustment set was $\{U_1, ..., U_{10}\}$ in Scenario~1, $\{U_1, ..., U_4\}$ in Scenarios~2, 2B and 3, $\{U_1, ..., U_5\}$ in Scenario~4 and 4B, and $\{U_1, U_2, U_3\}$ in Scenario~5. In real data analyses, the target adjustment set would typically be unknown. 

For all methods, only main terms of the covariates were considered. Since Scenarios 4, 4B and 5 feature non-linear and interaction terms, this allow evaluating the robustness of methods to model misspecifications. Most methods were implemented using the default options of their respective R function or package (see~Table \ref{table1}). For BAC, we used 500 burn-in iterations, followed by 5000 iterations with a thinning of 5. For GBCEE, we performed 2000 iterations of the Markov chain Monte Carlo model composition algorithm. 

For each method, we computed the bias as the difference between the average of the estimates and the true effect. In Scenarios 1-5, the true effect corresponded to the coefficient associated with $X$ in the data-generating equations. In Scenarios~2B and 4B, the true effect was estimated using Monte Carlo simulations, because the true risk difference cannot be easily determined analytically from the data-generating equations. A sample of $n$ = 1~000~000 observations was generated in which the counterfactual outcomes $Y^1$ and $Y^0$ were simulated for each observation, and the true effect was estimated as the mean difference of these counterfactual outcomes. The true effects were approximately 0.2814 and 0.0229 in Scenarios~2B and 4B, respectively. The standard deviation (SD) of the estimates, the mean squared error and the proportion of simulation replicates in which the 95\% confidence intervals included the true effect (CP) were also computed. The ratio of the root mean squared error of each method over the root mean squared error of target-g (Rel. RMSE) was also computed, except in Scenarios 4 and 4B where the comparator was target-AIPW, since the g-formula estimator was misspecified. For OAL, GLiDeR and MADR, confidence intervals were computed using 1000 nonparametric bootstrap replicates with the percentile method. Because this procedure is very computationally expansive, this was only done in Scenarios 4, 4B and 5. Additionally, we did not compute confidence intervals for the g-formula estimators in Scenarios 2B and 4B, since no simple variance estimator is available. For GBCEE and AIPW, confidence intervals were based on the efficient influence function variance estimator. The probability of inclusion of each covariate was calculated in Scenarios 1-5 for all methods, except HDM and CTMLE whose R function does not provide this information. 

\subsection{Results}

The results of the simulation study for Scenarios 1-5 and $n = 1000$ are reported in Table 2. The other results are overall similar and are presented in Web Appendix~B. Differences are noted in the main text as appropriate.    

First, we notice that the lowest bias, variance and RMSE is generally achieved by both the target-g and the target-AIPW, except in Scenario~4 where the g-formula performs poorly because of the outcome model misspecification. However, recall that these estimators are only considered as benchmarks. Second, we remark that the variance and RMSE of full-g and full-AIPW are greater than their target counterparts. The increase in the variance and RMSE is particularly pronounced for the double robust AIPW. These results are important to keep in mind when considering those for the variable selection methods: the double robustness property, which protects against bias due to model misspecification, may come at the cost of increased variance when spurious variables are included. 

Because BAC, BP and HDM are outcome-modeling based variable selection methods, it is not surprising that they generally perform better than the double robust variable selection methods (GBCEE, C-TMLE, GLiDeR and MADR) in Scenarios~1-3 and 5, where the outcome model is correctly specified. However, substantial bias is observed for BAC, BP and HDM in Scenario~4 where the outcome model is misspecified. Additionally, these methods offer no substantial RMSE reduction as compared to full-g when $n = 1000$, sometimes even performing worse. Some variance and RMSE reductions were however observed when $n = 200$ (see Web Appendix~B). OAL, which is an exposure-modeling based method, generally performed poorly as compared to all other variable selection methods, having larger bias and/or variance, except in Scenario~1. 

When comparing together the double robust methods, we first notice that, GBCEE, GLiDeR and MADR yield estimates with little or no bias in all scenarios when $n = 1000$, even when the outcome model or the exposure model is misspecified (Scenario~4 and 5, respectively). At most, a bias of -0.1 was observed for GLiDeR in Scenario~4. Unexpectedly, C-TMLE yields results with substantial bias in Scenario~4 where the outcome model is misspecified, although the bias is much smaller than that of the outcome-modeling methods (BAC, BP and HDM). When $n = 200$, larger biases were observed in Scenarios~3 and 4. In terms of RMSE, GBCEE, GLiDeR, MADR and C-TMLE all generally yield improvement improvement as compared to full-AIPW, except in Scenario~4 where the target adjustment set includes all covariates. In such a case, no variance reduction can be expected through variable selection. The relative performance of the methods is variable; no method is consistently performing better than all others across scenarios and sample sizes. For $n = 1000$, the two lowest RMSE among double robust variable selection methods are achieved, respectively, by GBCEE and MADR in Scenario~1; GLiDeR and C-TMLE in Scenario~2; GBCEE and MADR in Scenario~3; GBCEE and C-TMLE in Scenario~4; GBCEE and GLiDeR in Scenario~5.

The coverage probabilities of 95\% confidence intervals for BAC, BP and HDM are close to their nominal level in all scenarios, except Scenario~4 where the outcome-model is incorrectly specified. The coverage probability was much lower than 95\% in all scenarios for C-TMLE. For GBCEE, the coverage was close to 95\% in all scenarios, except in Scenario~5 where the coverage probability was 87\%. This is likely due to the inconsistency of the efficient influence function variance estimator when the exposure model is misspecified. To verify this assumption, we ran again the simulation for GBCEE using a nonparametric bootstrap variance estimator with 200 replicates. A 94\% coverage was then achieved. Similarly, in scenarios with $n = 200$, poorer coverage probabilities are obtained with GBCEE (see Web Appendix~B). This is most likely because of the asymptotic nature of the efficient influence function variance estimator, since poor coverage probabilities are also obtained with target-AIPW. As mentioned earlier, the coverage of OAL, GLiDeR and MADR were only examined in Scenarios~4 and 5 due to the associated computational burden. In Scenario~4, coverages of 89\%-90\% were observed, whereas the coverages were 93\%-95\% in Scenario~5.

Plots of the inclusion probability of covariates are available in Web Appendix~C. These figures reveal that GBCEE, OAL, GLiDeR and MADR have relatively similar behaviors, having large probabilities of including both true confounders and outcome risk factors, and lower probabilities of including the other variables. BAC and BP have high probabilities of including true confounders, outcome risk factors and instruments.

We evaluated the computational time of the variable selection methods in one replication of Scenario~1 with $n = 1000$ on a PC with 4 GHz and 16 Gb RAM. The running time for producing both the estimate and confidence interval was 25.8 seconds for GBCEE using the efficient influence function variance estimator and 40.2 minutes when using 200 bootstrap replicates for estimating the variance, 1 minute for C-TMLE, 1.6 minutes for BAC, 0.6 seconds for BP and 2 seconds for HDM. The execution time for obtaining point estimates only was 0.7 second for OAL, 8.7 seconds for GLiDeR, and 1.1 minutes for MADR. Obtaining the confidence interval required 8.7 minutes for OAL, 39.8 minutes for GLiDeR and 19.9 hours for MADR, using 1000 bootstrap replicates.

\begin{table}[!htbp] \centering 
  \caption{Results of Scenarios 1-5, with $n$ = 1000} 
\resizebox{\textwidth}{!}{	
\begin{tabular}{ccccc|cccc|cccc} 
\\[-1.8ex]\hline 
& \multicolumn{4}{c}{ Scenario 1} & \multicolumn{4}{c}{ Scenario 2} & \multicolumn{4}{c}{ Scenario 3}\\
\multicolumn{1}{c}{} & \multicolumn{1}{c}{} & \multicolumn{1}{c}{} & \multicolumn{1}{c}{Rel.} & \multicolumn{1}{c}{}
 & \multicolumn{1}{c}{} & \multicolumn{1}{c}{} & \multicolumn{1}{c}{Rel.} & \multicolumn{1}{c}{}
 & \multicolumn{1}{c}{} & \multicolumn{1}{c}{} & \multicolumn{1}{c}{Rel.} & \multicolumn{1}{c}{} \\ 
\multicolumn{1}{c}{Method} & \multicolumn{1}{c}{Bias} & \multicolumn{1}{c}{SD} & \multicolumn{1}{c}{RMSE} & \multicolumn{1}{c}{CP}
 & \multicolumn{1}{c}{Bias} & \multicolumn{1}{c}{SD} & \multicolumn{1}{c}{RMSE} & \multicolumn{1}{c}{CP}
 & \multicolumn{1}{c}{Bias} & \multicolumn{1}{c}{SD} & \multicolumn{1}{c}{RMSE} & \multicolumn{1}{c}{CP} \\ 
\hline \\[-1.8ex] 
\multicolumn{1}{l}{full-g}         & 0.00 & 0.09 & 1.41 & 0.96    & 0.00 & 0.09 & 1.06 & 0.95  & 0.01 & 0.18 & 1.14 & 0.95 \\
\multicolumn{1}{l}{target-g}       & 0.00 & 0.06 & 1.00 & 0.96    & 0.00 & 0.08 & 1.00 & 0.95  & 0.00 & 0.16 & 1.00 & 0.95 \\
\multicolumn{1}{l}{full-AIPW}   & 0.03 & 0.65 & 10.11 & 0.94   & -0.01 & 0.45 & 5.40 & 0.94 & 0.00 & 0.98 & 6.12 & 0.95 \\
\multicolumn{1}{l}{target-AIPW} & 0.00 & 0.06 & 1.00 & 0.95    & 0.00 & 0.08 & 1.00 & 0.86  & 0.00 & 0.16 & 1.00 & 0.89 \\
\multicolumn{1}{l}{GBCEE}        & 0.00 & 0.08 & 1.17 & 0.95    & 0.00 & 0.14 & 1.65 & 0.94  & 0.00 & 0.20 & 1.29 & 0.94 \\
\multicolumn{1}{l}{OAL}          & 0.02 & 0.08 & 1.22 & .       & 0.26 & 0.25 & 4.29 & .     & -0.36 & 0.29 & 2.88 & .   \\
\multicolumn{1}{l}{C-TMLE}       & 0.01 & 0.10 & 1.50 & 0.87    & 0.02 & 0.11 & 1.39 & 0.77  & -0.00 & 0.28 & 1.78 & 0.65 \\
\multicolumn{1}{l}{GLiDeR}       & 0.01 & 0.08 & 1.18 & .       & 0.00 & 0.12 & 1.38 & .     & -0.10 & 0.21 & 1.45 & .   \\
\multicolumn{1}{l}{BAC}          & 0.00 & 0.09 & 1.42 & 0.96   & 0.00 & 0.09 & 1.06 & 0.95 & 0.00 & 0.18 & 1.11 & 0.92 \\
\multicolumn{1}{l}{MADR}         & 0.01 & 0.07 & 1.05 & .       & 0.00 & 0.15 & 1.75 & .     & 0.00 & 0.22 & 1.41 & .    \\
\multicolumn{1}{l}{BP}           & 0.00 & 0.09 & 1.42 & 0.95    & 0.00 & 0.09 & 1.06 & 0.95  & 0.00 & 0.18 & 1.13 & 0.94 \\
\multicolumn{1}{l}{HDM}          & 0.00 & 0.09 & 1.41 & 0.96   & 0.00 & 0.09 & 1.06 & 0.95  & 0.00 & 0.18 & 1.10 & 0.94 \\
\hline \\[-1.8ex] 
& \multicolumn{4}{c}{ Scenario 4}& \multicolumn{4}{c}{ Scenario 5}  & \multicolumn{4}{c}{}\\
\hline \\[-1.8ex] 
\multicolumn{1}{l}{full-g}          & -4.63 & 1.67 & 1.96 & 0.26  & 0.00 & 0.10 & 1.02 & 0.95 & \multicolumn{4}{c}{}\\ 
\multicolumn{1}{l}{target-g}        & -4.63 & 1.67 & 1.96 & 0.26  & 0.00 & 0.09 & 1.00 & 0.95 & \multicolumn{4}{c}{}\\ 
\multicolumn{1}{l}{full-AIPW}     & -0.01 & 2.51 & 1.00 & 0.93  & 0.00 & 0.13 & 1.41 & 0.92 & \multicolumn{4}{c}{}\\  
\multicolumn{1}{l}{target-AIPW}   & -0.01 & 2.51 & 1.00 & 0.93  & 0.00 & 0.09 & 1.00 & 0.76 & \multicolumn{4}{c}{}\\ 
\multicolumn{1}{l}{GBCEE}         & 0.07 & 2.51 & 1.00 & 0.91   & 0.00 & 0.11 & 1.20 & 0.88 & \multicolumn{4}{c}{}\\ 
\multicolumn{1}{l}{OAL}           & 0.14 & 6.20 & 2.47 & 0.89   & 0.15 & 0.22 & 2.83 & 0.93 & \multicolumn{4}{c}{}\\ 
\multicolumn{1}{l}{C-TMLE}        & 2.45 & 1.10 & 1.07 & 0.39   & 0.02 & 0.17 & 1.78 & 0.72 & \multicolumn{4}{c}{}\\ 
\multicolumn{1}{l}{GLiDeR}        & 0.01 & 3.34 & 1.33 & 0.89   & 0.02 & 0.13 & 1.41 & 0.95 & \multicolumn{4}{c}{}\\ 
\multicolumn{1}{l}{BAC}           & -4.66 & 1.67 & 1.97 & 0.24  & 0.00 & 0.10 & 1.02 & 0.95 & \multicolumn{4}{c}{}\\  
\multicolumn{1}{l}{MADR}          & 0.02 & 3.22 & 1.28 & 0.90   & 0.03 & 0.26 & 2.82 & 0.95 & \multicolumn{4}{c}{}\\ 
\multicolumn{1}{l}{BP}            & -4.63 & 1.67 & 1.96 & 0.26  & 0.00 & 0.10 & 1.02 & 0.95 & \multicolumn{4}{c}{}\\ 
\multicolumn{1}{l}{HDM}           & -4.63 & 1.67 & 1.96 & 0.20  & 0.00 & 0.10 & 1.02 & 0.95 & \multicolumn{4}{c}{}\\ 
\hline \\[-1.8ex] 
\end{tabular}}
\end{table} 

\section{Application for estimating the effect of physical activity on the risk of fractures} \label{application}

Osteoporosis is a disease where bones become fragile because of low bone mineral density or to the deterioration of bone architecture. It is a common disease with a prevalence of 5\% in men and 25\% in women aged 65 years and older from the United States \cite{Looker}. The prevalence of osteoporosis increases with age. Because of the bone fragility, individuals with osteoporosis are at higher risk of fractures. In addition to important pain, osteoporotic fractures may also induce a loss of mobility and quality of life, additional morbidities, and premature mortality. Preventing osteoporosis and the related fractures has thus become a public health priority. Physical activity may help preventing osteoporosis and osteoporotic fractures. Indeed, regular physical activity is associated with increased bone mineral density and lower risk of falls \cite{Bonaiuti, Thibaud}. 

To illustrate the use of GBCEE, we estimated the effect of attaining the physical activity recommendations from the World Health Organization on the 5-year risk of fracture in elderly women using the publicly available limited data set from the Study of  Osteoporotic Fractures \cite{WHO}. The Study of Osteoporotic Fractures is a multicentric population based cohort study that recruited women aged 65 or older in four urban regions in the United states: Baltimore, Pittsburgh, Minneapolis, and Portland. More details can be found elsewhere. An ethical exemption was obtained from the CHU de Qu\'ebec -- Universit\'e Laval Ethics Board (\# 2020-4788). 

In this application, we consider data on 9671 white women enrolled in 1986. Subjects were considered as exposed if they spent 7.5 kilocalories per kilogram of body mass from moderate or high intensity physical activity per week, and unexposed otherwise. The outcome of interest was the occurrence of any hip or upper leg fracture in the five years following the baseline interview. A very rich set of 55 measured potential confounders were identified based on substantive knowledge and notably include data on age, ethnic origin, body mass index, smoking, alcohol use, education, various drug use, fall history, self and familial history of fracture, physical activity history when a teenager and during adulthood, fear of falling and various health conditions (see Web Appendix~D).

Table~8 in Web Appendix~D presents a comparison of the baseline characteristics of participants according to exposure to physical activity recommendations. Among others, subjects that were unexposed had a greater body mass index, were older, had less years of education, were more afraid of falling, practiced less physical activity in the past, drank less alcohol, had more difficulty walking, rated their overall health more poorly, and were less likely to still be married. 

Only 3,458 participants had complete information for all variables. Most variables had only few missing data ($<$10\%), except for parental history of fracture (mother = 24\%, father = 37\%) and years since menopause (18\%). Missing data were imputed using chained equations with the \texttt{mice} package in R. To simplify this illustration, a single random imputation was performed, but multiple imputations would be preferable in practice. The effect was estimated with AIPW adjusting for all potential confounders and GBCEE. Standard errors were estimated using 50 nonparametric bootstrap replicates. 

The fully adjusted AIPW yielded a 5-year risk difference of fracture of -1.1\% (95\% confidence interval: -4.6\%, 2.4\%) between subjects that attained the physical activity recommendations and those who did not. The estimate using GBCEE, with $\omega = 500\sqrt{n}$ and 20~000 iterations, was -0.9\% (95\% confidence interval: -2.3\%, 0.4\%). In this illustration, GBCEE yields a major decrease of the width of the 95\% confidence interval as compared to the fully adjusted estimate (61\%). The potential confounders with the largest probability of being selected by GBCEE were: mother history of fracture (100\%), self-rated health (100\%), difficulty to walk 2-3 blocs (100\%), the presence of fracture before age 50 (100\%), education (100\%), age (100\%), body mass index (100\%), short mini mental status exam (80\%), any drinking in past year (73\%), and osteoporosis diagnostic (72\%). All other covariates had a probability of being selected $<$ 50\%, with most being null. 

\section{Discussion}
\label{s:discuss}

We have presented a generalization of the Bayesian Causal Effect Estimation algorithm to estimate the causal effect of a binary or continuous exposure on a binary or continuous outcome using double robust targeted maximum likelihood estimators. Like the original BCEE, GBCEE is theoretically motivated using the graphical framework to causal inference. It aims to select adjustment covariates such that the final estimator is unbiased and has reduced variance as compared to an estimator adjusting for all covariates. Additionally, the Bayesian framework allows producing inferences that account for the model selection step in a principled way. We have also proposed an implementation of GBCEE that is more computationally efficient than that of BCEE.

We have compared GBCEE to alternative model selection methods in a simulation study. GBCEE produced estimates with little or no bias in all scenarios, and with improved precision as compared to a fully adjusted double robust estimator in most scenarios. The only situation where no improvement in precision was observed was when unbiased estimation required adjusting for all potential confounders; hence no variance reduction was possible. The relative performance of the estimators was observed to depend both on the data-generating mechanism and sample size, but GBCEE often had the lowest or second lowest RMSE among double robust methods. It might be interesting for future studies to specifically investigate the factors that affect the efficiency of the different model selection methods and to provide insights regarding the situations where one method should be expected to outperform the others. The simulation study results also indicated that outcome-model based algorithms outperformed double robust methods in scenarios where the outcome model was correctly specified. However, the former methods produced estimates with high bias when the outcome model was incorrectly specified, unlike the latter. Model misspecifications are likely to be common in practice. 

From a computational perspective, GBCEE produced both an estimate and inferences more quickly than the other double robust algorithms when the influence function based variance estimator was used. This variance estimator produced valid inferences in most scenarios. Two exceptions were when the exposure model was misspecified and when the sample size was small. As such, it is advisable to use the bootstrap variance estimator for inferences when the sample size is moderate or small. In such situations, using the bootstrap variance estimator is unlikely to be computationally prohibitive. 

Regarding inferences for model selection procedures, it is unclear if employing the usual nonparametric bootstrap is appropriate. Indeed, model selection may yield estimators that lack the smoothness required for the bootstrap to be valid. This may be the reason why the coverage of 95\% confidence intervals was around 90\% for MADR, GLiDeR and OAL in Scenario~4 of our simulation study. A smoothed-bootstrap method has been proposed specifically for the variable selection problem, \cite{Efron} and was used by Shortreed and Erterfaie (2017)\cite{Shortreed}. We have explored using this approach in our simulation study, but we have observed that a very large number of replications (5000-10~000) was necessary to achieve appropriate variance estimation. Shortreed and Ertefaie (2017) \cite{Shortreed} based their inferences on 10~000 replications. Overall, we believe that inference procedures for variable selection methods require further investigation. We note that GBCEE does not share this limitation when the bootstrap variance estimator is used, since the bootstrap is performed within -- and not over -- the variable selection algorithm. 

Finally, we have illustrated the use of GBCEE for estimating the effect of reaching physical activity recommendations on the risk of osteoporotic fractures among elderly women. In this example, the estimate produced by GBCEE was similar to that of the fully adjusted double robust estimator, but the confidence interval was much shorter. Since resources available for research are limited, it is important to make the most out of the available data. Confounder selection methods should be considered as a valuable tool for achieving this, especially when there are multiple potential confounders or when it is unknown whether some covariates are confounders, risk factors for the outcome, or instruments.

There are multiple directions in which the current work could be extended. For instance, to the best of our knowledge, no confounder selection method is applicable to the censored time-to-event outcome case yet. To reduce the risk of model misspecifications, it would also be possible to employ machine learning within GBCEE. In fact, TMLE is often paired with Super Learner, an ensemble method that combines the predictions from multiple methods using a cross-validation procedure \cite{vanderLaan2011}. A simple possibility for doing so would entail computing the prior and posterior probability of a given adjustment set $\bm{\alpha^Y}$ using parametric working models, exactly as proposed in the current paper. However, the causal contrast estimate $\hat{\Delta}^{\bm{\alpha^Y}}$ would be computed using a TMLE that employs machine learning. We would expect this procedure to perform well in practice as long as confounders feature a non negligible main linear effect on both the exposure and the outcome. However, an important downside to such a procedure would be its computational burden.

\section*{Acknowledgements}

This work was supported by grants from the Natural Sciences and Engineering Research Council of Canada and the Fonds de recherche du Qu\'ebec – Sant\'e. Dr Talbot is a Fonds de recherche du Qu\'ebec – Sant\'e Chercheur-Boursier. The Study of Osteoporotic Fractures (SOF) is supported by National Institutes of Health funding. The National Institute on Aging (NIA) provides support under the following grant numbers: R01 AG005407, R01 AR35582, R01 AR35583, R01 AR35584, R01 AG005394, R01 AG027574, and R01 AG027576. 

\clearpage

\section*{Web Appendix A - Detailed presentation of the double-robust estimators}

\subsection*{Continuous outcome, continuous exposure}

The causal contrast of interest is $\Delta = \mathbb{E}[Y^{x+1}] - \mathbb{E}[Y^{x}]$. The algorithm for obtaining the TMLE is
\begin{align*}
\hat{\epsilon} &= \underset{\epsilon}{argmin} \sum_{i = 1}^n \left[y_i - \hat{y}_i^{\bm{\alpha^Y}} - \epsilon \left(x - \hat{x}_i^{\bm{\alpha^Y}}\right) \right]^2 \\
\hat{\Delta}^{\bm{\alpha^Y}} &= \beta^{\bm{\alpha^Y}} + \hat{\epsilon},
\end{align*}
\noindent where $\hat{y}_i^{\bm{\alpha^Y}}$ and $\hat{x}_i^{\bm{\alpha^Y}}$ are the predicted values from the outcome and the exposure models, respectively. Recall that each model include the covariates $\bm{\alpha^Y}$. In practice, $\hat{\epsilon}$ can be obtained as the coefficient estimate of a linear regression without intercept of $Y$ on $x - \hat{x}_i^{\bm{\alpha^Y}}$ with $\hat{y}_i^{\bm{\alpha^Y}}$ as an offset term.

\subsection*{Continuous outcome, binary exposure\label{conty_binx}} 
The causal contrast of interest is $\Delta = \mathbb{E}[Y^1] - \mathbb{E}[Y^0]$. The TMLE of $\mathbb{E}[Y^x]$, for $x = 0, 1$, is obtained as follows:
\begin{align*}
&w_i = \frac{I(X_i = x)}{\hat{P}(X = x|\bm{\alpha^Y}, \bm{U}_i)}, \ \ Q^x_{i0} = \hat{\mathbb{E}}[Y|\bm{\alpha^Y}, X = x, \bm{U}_i], \\  
&\hat{\epsilon}^x = \underset{\epsilon^x}{argmin} \sum_{i = 1}^n w_i \left[y_i - Q^x_{i0} - \epsilon^x \right]^2, \ \ Q^x_{i1} = Q^x_{i0} + \hat{\epsilon}^x, \\
&\hat{\mathbb{E}}[Y^x] = \frac{\sum_{i=1}^n Q^x_{i1}}{n},
\end{align*}

\noindent where $I(\cdot)$ is the indicator function taking value 1 when its argument is true and 0 otherwise, and $\hat{P}(X = 1|\bm{\alpha^Y}, \bm{U}_i)$ and $\hat{\mathbb{E}}[Y|\bm{\alpha^Y}, X = x, \bm{U}_i]$ are calculated from (2) and (1) in the main manuscript, respectively. The intercept estimate of a linear regression of $Y$ with only an intercept and offset term $Q^x_{i0}$ weighted according to $w_i$ can be used to obtain $\hat{\epsilon}$. Then, $\hat{\Delta}^{\bm{\alpha^Y}} = \hat{\mathbb{E}}[Y^1] - \hat{\mathbb{E}}[Y^0]$.

\subsection*{Binary outcome, binary exposure}

The same algorithm as in Section \ref{conty_binx} is employed for estimating $\mathbb{E}[Y^x]$. 

\subsection*{Binary outcome, continuous exposure}

The contrasts we consider are $\Delta = \mathbb{E}[Y^x] - \mathbb{E}[Y^{x'}]$ or $\Delta = \mathbb{E}[Y^{x}]/\mathbb{E}[Y^{x'}]$, with $x \neq x'$. The algorithm for estimating $\Delta^{\bm{\alpha^Y}}$ is:
\begin{align*}
Q^x_{i0} &= \hat{\mathbb{E}}[Y|\bm{\alpha^Y}, X = x, \bm{U}_i], \\
\hat{\epsilon} &= \underset{\epsilon}{argmax} \ expit\left(logit(Q^x_{i0}) + \epsilon \frac{f_X(x)}{f_X(x|\bm{\alpha^Y}, \bm{U}_i)} \right)^Y \times \\
&  \left[1 - expit\left(logit(Q^x_{i0}) + \epsilon \frac{f_X(x)}{f_X(x|\bm{\alpha^Y}, \bm{U}_i)} \right)\right]^{1-Y}, \\
Q^x_{i1} &= expit\left[logit(Q^x_{i0}) + \hat{\epsilon}\frac{f_X(x)}{f_X(x|\bm{\alpha^Y}, \bm{U}_i)}\right], \\
\hat{\mathbb{E}}[Y^x] &= \frac{\sum_{i=1}^n Q^x_{i1}}{n},
\end{align*}
\noindent where $f_X(x)$ is the marginal density of $X$ assuming a normal density. The quantity $\hat{\epsilon}$ can be conveniently estimated as the unique coefficient of a logistic regression of $Y$ on $f_X(x)/f_X(x|\bm{\alpha^Y}, \bm{U}_i)$ with no intercept and offset term $logit(Q^x_{i0})$. 

\clearpage
 
\section*{Web Appendix B - Additional simulation results}

\begin{table}[!htbp] \centering 
  \caption{Results of Scenario 1, with $n$ = 200} 
\begin{tabular}{@{\extracolsep{5pt}} D{.}{.}{-2} D{.}{.}{-2} D{.}{.}{-2} D{.}{.}{-2} D{.}{.}{-2} } 
\\[-1.8ex] \hline 
\hline \\[-1.8ex] 
\multicolumn{1}{c}{} & \multicolumn{1}{c}{} & \multicolumn{1}{c}{} & \multicolumn{1}{c}{Rel.} & \multicolumn{1}{c}{} \\ 
\multicolumn{1}{c}{Method} & \multicolumn{1}{c}{Bias} & \multicolumn{1}{c}{SD} & \multicolumn{1}{c}{RMSE} & \multicolumn{1}{c}{CP} \\ 
\hline \\[-1.8ex] 
\multicolumn{1}{l}{full-g} & -0.00 & 0.24 & 1.49 & 0.95 \\ 
\multicolumn{1}{l}{target-g} & 0.00 & 0.16 & 1.00 & 0.95 \\ 
\multicolumn{1}{l}{full-AIPW} & 1.13 & 31.36 & 198.12 & 0.82 \\ 
\multicolumn{1}{l}{target-AIPW} & 0.00 & 0.16 & 1.00 & 0.93 \\ 
\multicolumn{1}{l}{GBCEE} & 0.02 & 0.20 & 1.29 & 0.93 \\ 
\multicolumn{1}{l}{OAL} & 0.05 & 0.20 & 1.31 & . \\ 
\multicolumn{1}{l}{C-TMLE} & 0.06 & 0.28 & 1.82 & 0.74 \\ 
\multicolumn{1}{l}{GLiDeR} & 0.03 & 0.18 & 1.13 & . \\ 
\multicolumn{1}{l}{BAC} & 0.01 & 0.22 & 1.38 & 0.94 \\ 
\multicolumn{1}{l}{MADR} & 0.03 & 0.17 & 1.12 & . \\ 
\multicolumn{1}{l}{BP} & 0.02 & 0.22 & 1.42 & 0.92 \\ 
\multicolumn{1}{l}{HDM} & -0.00 & 0.18 & 1.14 & 0.95 \\ 
\hline \\[-1.8ex] 
\end{tabular} 
\end{table} 

\begin{table}[!htbp] \centering 
  \caption{Results of Scenario 2, with $n$ = 200} 
\begin{tabular}{@{\extracolsep{5pt}} D{.}{.}{-2} D{.}{.}{-2} D{.}{.}{-2} D{.}{.}{-2} D{.}{.}{-2} } 
\\[-1.8ex]\hline 
\hline \\[-1.8ex] 
\multicolumn{1}{c}{} & \multicolumn{1}{c}{} & \multicolumn{1}{c}{} & \multicolumn{1}{c}{Rel.} & \multicolumn{1}{c}{} \\ 
\multicolumn{1}{c}{Method} & \multicolumn{1}{c}{Bias} & \multicolumn{1}{c}{SD} & \multicolumn{1}{c}{RMSE} & \multicolumn{1}{c}{CP} \\ 
\hline \\[-1.8ex] 
\multicolumn{1}{l}{full-g} & -0.01 & 0.21 & 1.10 & 0.95 \\ 
\multicolumn{1}{l}{target-g} & -0.01 & 0.19 & 1.00 & 0.95 \\ 
\multicolumn{1}{l}{full-AIPW} & -0.02 & 2.27 & 12.01 & 0.88 \\ 
\multicolumn{1}{l}{target-AIPW} & -0.01 & 0.19 & 1.00 & 0.85 \\ 
\multicolumn{1}{l}{GBCEE} & -0.01 & 0.26 & 1.39 & 0.91 \\ 
\multicolumn{1}{l}{OAL} & 0.41 & 0.46 & 3.27 & . \\ 
\multicolumn{1}{l}{C-TMLE} & 0.05 & 0.28 & 1.48 & 0.70 \\ 
\multicolumn{1}{l}{GLiDeR} & 0.01 & 0.24 & 1.25 & . \\ 
\multicolumn{1}{l}{BAC} & -0.04 & 0.20 & 1.09 & 0.94 \\ 
\multicolumn{1}{l}{MADR} & -0.01 & 0.40 & 2.13 & . \\ 
\multicolumn{1}{l}{BP} & -0.01 & 0.21 & 1.10 & 0.94 \\ 
\multicolumn{1}{l}{HDM} & -0.00 & 0.20 & 1.07 & 0.95 \\ 
\hline \\[-1.8ex] 
\end{tabular} 
\end{table} 

\begin{table}[!htbp] \centering 
  \caption{Results of Scenario 3, with $n$ = 200} 
  \begin{tabular}{@{\extracolsep{5pt}} D{.}{.}{-2} D{.}{.}{-2} D{.}{.}{-2} D{.}{.}{-2} D{.}{.}{-2} } 
\\[-1.8ex]\hline 
\hline \\[-1.8ex] 
\multicolumn{1}{c}{} & \multicolumn{1}{c}{} & \multicolumn{1}{c}{} & \multicolumn{1}{c}{Rel.} & \multicolumn{1}{c}{} \\ 
\multicolumn{1}{c}{Method} & \multicolumn{1}{c}{Bias} & \multicolumn{1}{c}{SD} & \multicolumn{1}{c}{RMSE} & \multicolumn{1}{c}{CP} \\ 
\hline \\[-1.8ex] 
\multicolumn{1}{l}{full-g} & 0.00 & 0.54 & 1.52 & 0.95 \\ 
\multicolumn{1}{l}{target-g} & -0.01 & 0.35 & 1.00 & 0.94 \\ 
\multicolumn{1}{l}{full-AIPW} & . & . & . & . \\ 
\multicolumn{1}{l}{target-AIPW} & -0.01 & 0.35 & 1.00 & 0.88 \\ 
\multicolumn{1}{l}{GBCEE} & -0.13 & 0.48 & 1.41 & 0.79 \\ 
\multicolumn{1}{l}{OAL} & -0.37 & 0.52 & 1.79 & . \\ 
\multicolumn{1}{l}{C-TMLE} & 0.02 & 1.08 & 3.05 & 0.56 \\ 
\multicolumn{1}{l}{GLiDeR} & -0.16 & 0.38 & 1.18 & . \\ 
\multicolumn{1}{l}{BAC} & -0.04 & 0.47 & 1.32 & 0.94 \\ 
\multicolumn{1}{l}{MADR} & -0.14 & 0.48 & 1.41 & . \\ 
\multicolumn{1}{l}{BP} & 0.01 & 0.45 & 1.28 & 0.93 \\ 
\multicolumn{1}{l}{HDM} & -0.01 & 0.36 & 1.01 & 0.95 \\ 
\hline \\[-1.8ex] 
\end{tabular} 
\end{table}

\begin{table}[!htbp] \centering 
  \caption{Results of Scenario 4, with $n$ = 200} 
   
\begin{tabular}{@{\extracolsep{5pt}} D{.}{.}{-2} D{.}{.}{-2} D{.}{.}{-2} D{.}{.}{-2} D{.}{.}{-2} } 
\\[-1.8ex]\hline 
\hline \\[-1.8ex] 
\multicolumn{1}{c}{} & \multicolumn{1}{c}{} & \multicolumn{1}{c}{} & \multicolumn{1}{c}{Rel.} & \multicolumn{1}{c}{} \\ 
\multicolumn{1}{c}{Method} & \multicolumn{1}{c}{Bias} & \multicolumn{1}{c}{SD} & \multicolumn{1}{c}{RMSE} & \multicolumn{1}{c}{CP} \\ 
\hline \\[-1.8ex] 
\multicolumn{1}{l}{full/target-g} & -4.78 & 3.82 & 1.15 & 0.83 \\ 
\multicolumn{1}{l}{full/target-AIPW} & 0.23 & 5.30 & 1.00 & 0.94 \\ 
\multicolumn{1}{l}{GBCEE} & 0.47 & 5.36 & 1.01 & 0.92 \\ 
\multicolumn{1}{l}{OAL} & 1.44 & 11.37 & 2.16 & 0.83 \\ 
\multicolumn{1}{l}{C-TMLE} & 2.50 & 2.79 & 0.71 & 0.78 \\ 
\multicolumn{1}{l}{GLiDeR} & 0.21 & 5.43 & 1.02 & 0.90 \\ 
\multicolumn{1}{l}{BAC} & -4.61 & 3.82 & 1.13 & 0.84 \\ 
\multicolumn{1}{l}{MADR} & 0.28 & 5.13 & 0.97 & 0.90 \\ 
\multicolumn{1}{l}{BP} & -4.78 & 3.81 & 1.15 & 0.83 \\ 
\multicolumn{1}{l}{HDM} & -4.77 & 3.82 & 1.15 & 0.76 \\ 
\hline \\[-1.8ex] 
\end{tabular} 
\end{table} 

\begin{table}[!htbp] \centering 
  \caption{Results for Scenario 5, with $n$ = 200} 
   
\begin{tabular}{@{\extracolsep{5pt}} D{.}{.}{-2} D{.}{.}{-2} D{.}{.}{-2} D{.}{.}{-2} D{.}{.}{-2} } 
\\[-1.8ex]\hline 
\hline \\[-1.8ex] 
\multicolumn{1}{c}{} & \multicolumn{1}{c}{} & \multicolumn{1}{c}{} & \multicolumn{1}{c}{Rel.} & \multicolumn{1}{c}{} \\ 
\multicolumn{1}{c}{Method} & \multicolumn{1}{c}{Bias} & \multicolumn{1}{c}{SD} & \multicolumn{1}{c}{RMSE} & \multicolumn{1}{c}{CP} \\ 
\hline \\[-1.8ex] 
\multicolumn{1}{l}{full-g} & -0.01 & 0.22 & 1.03 & 0.95 \\ 
\multicolumn{1}{l}{target-g} & -0.00 & 0.21 & 1.00 & 0.95 \\ 
\multicolumn{1}{l}{full-AIPW} & -0.06 & 1.53 & 7.14 & 0.86 \\ 
\multicolumn{1}{l}{target-AIPW} & -0.00 & 0.21 & 1.00 & 0.73 \\ 
\multicolumn{1}{l}{GBCEE} & 0.01 & 0.26 & 1.23 & 0.85 \\ 
\multicolumn{1}{l}{OAL} & 0.26 & 0.42 & 2.29 & 0.91 \\ 
\multicolumn{1}{l}{C-TMLE} & 0.05 & 0.40 & 1.87 & 0.66 \\ 
\multicolumn{1}{l}{GLiDeR} & 0.04 & 0.23 & 1.10 & 0.96 \\ 
\multicolumn{1}{l}{BAC} & 0.01 & 0.22 & 1.02 & 0.95 \\ 
\multicolumn{1}{l}{MADR} & 0.01 & 0.65 & 3.04 & 0.97 \\ 
\multicolumn{1}{l}{BP} & -0.01 & 0.22 & 1.03 & 0.95 \\ 
\multicolumn{1}{l}{HDM} & -0.01 & 0.22 & 1.03 & 0.95 \\ 
\hline \\[-1.8ex] 
\end{tabular} 
\end{table} 
\vspace*{-8pt}

\begin{table}[!htbp] \centering 
  \caption{Results of Scenario 2B, with $n$ = 1000} 
   
\begin{tabular}{@{\extracolsep{5pt}} D{.}{.}{-2} D{.}{.}{-2} D{.}{.}{-2} D{.}{.}{-2} D{.}{.}{-2} } 
\\[-1.8ex]\hline 
\hline \\[-1.8ex] 
\multicolumn{1}{c}{} & \multicolumn{1}{c}{} & \multicolumn{1}{c}{} & \multicolumn{1}{c}{Rel.} & \multicolumn{1}{c}{} \\ 
\multicolumn{1}{c}{Method} & \multicolumn{1}{c}{Bias} & \multicolumn{1}{c}{SD} & \multicolumn{1}{c}{RMSE} & \multicolumn{1}{c}{CP} \\ 
\hline \\[-1.8ex] 
\multicolumn{1}{l}{full-g} & 0.00 & 0.04 & 1.11 & . \\ 
\multicolumn{1}{l}{target-g} & 0.00 & 0.04 & 1.00 & . \\ 
\multicolumn{1}{l}{full-AIPW} & -0.00 & 0.09 & 2.67 & 0.93 \\ 
\multicolumn{1}{l}{target-AIPW} & 0.00 & 0.05 & 1.33 & 0.93 \\ 
\multicolumn{1}{l}{GBCEE} & 0.00 & 0.05 & 1.54 & 0.93 \\ 
\multicolumn{1}{l}{OAL} & 0.04 & 0.07 & 2.17 & . \\ 
\multicolumn{1}{l}{C-TMLE} & 0.04 & 0.08 & 2.39 & 0.53 \\ 
\multicolumn{1}{l}{BAC} & -0.01 & 0.04 & 1.10 & 0.91 \\ 
\hline \\[-1.8ex] 
\end{tabular} 
\end{table} 

\begin{table}[!htbp] \centering 
  \caption{Results of Scenario 4B, with $n$ = 1000} 
   
\begin{tabular}{@{\extracolsep{5pt}} D{.}{.}{-2} D{.}{.}{-2} D{.}{.}{-2} D{.}{.}{-2} D{.}{.}{-2} } 
\\[-1.8ex]\hline 
\hline \\[-1.8ex] 
\multicolumn{1}{c}{} & \multicolumn{1}{c}{} & \multicolumn{1}{c}{} & \multicolumn{1}{c}{Rel.} & \multicolumn{1}{c}{} \\ 
\multicolumn{1}{c}{Method} & \multicolumn{1}{c}{Bias} & \multicolumn{1}{c}{SD} & \multicolumn{1}{c}{RMSE} & \multicolumn{1}{c}{CP} \\ 
\hline \\[-1.8ex] 
\multicolumn{1}{l}{full/target-g} & 0.00 & 0.02 & 1.02 & . \\ 
\multicolumn{1}{l}{full/target-AIPW} & -0.00 & 0.02 & 1.00 & 0.94 \\ 
\multicolumn{1}{l}{GBCEE} & -0.00 & 0.02 & 1.00 & 0.94 \\ 
\multicolumn{1}{l}{OAL} & 0.00 & 0.02 & 1.06 & 0.96 \\ 
\multicolumn{1}{l}{C-TMLE} & 0.00 & 0.02 & 1.00 & 0.96 \\ 
\multicolumn{1}{l}{BAC} & 0.00 & 0.02 & 1.02 & 0.95 \\ 
\hline \\[-1.8ex] 
\end{tabular} 
\end{table} 

\clearpage

\section*{Web Appendix C - Covariates inclusion probabilities}

In the plots below, MADRx and MADRy represent the covariates included to model the exposure and the outcome, respectively. Unlike GBCEE,  MADR does not necessarily include the same covariates in both models. 

\includegraphics{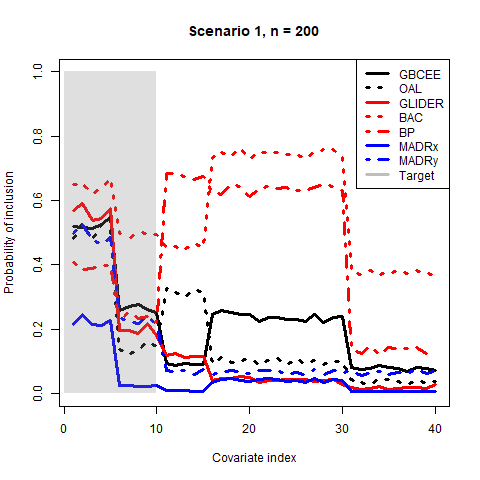}
\clearpage
\includegraphics{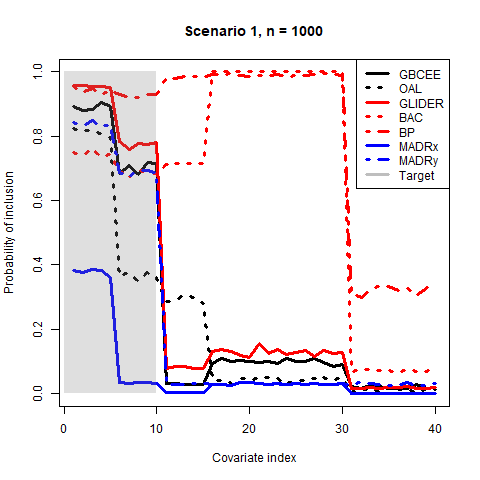}
\clearpage
\includegraphics{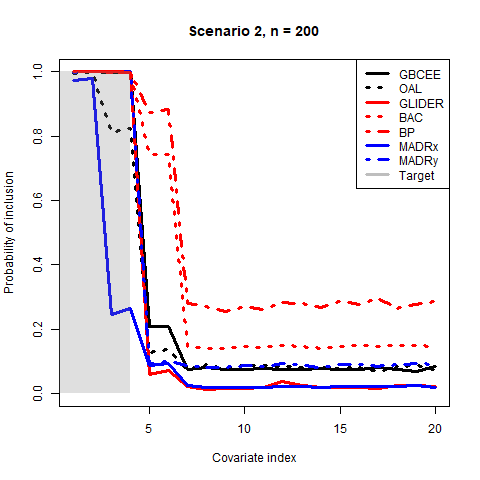}
\clearpage
\includegraphics{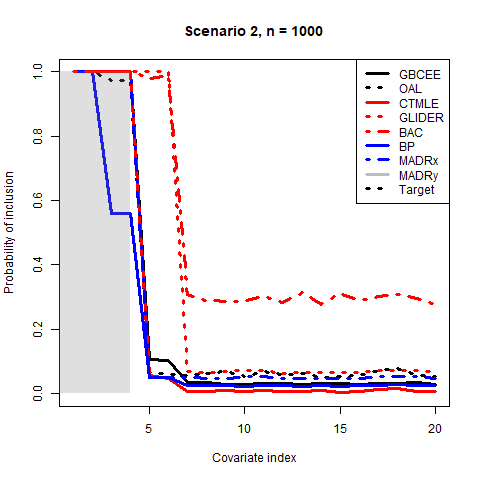}
\clearpage
\includegraphics{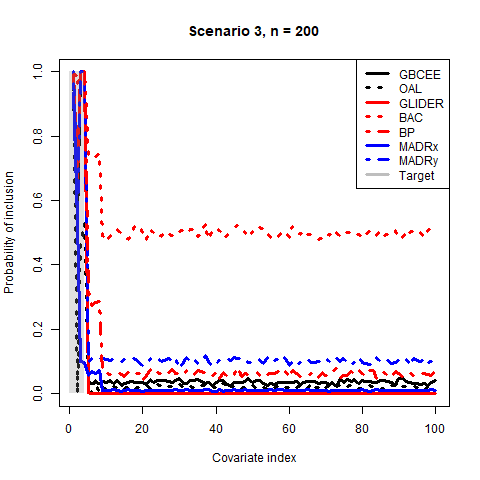}
\clearpage
\includegraphics{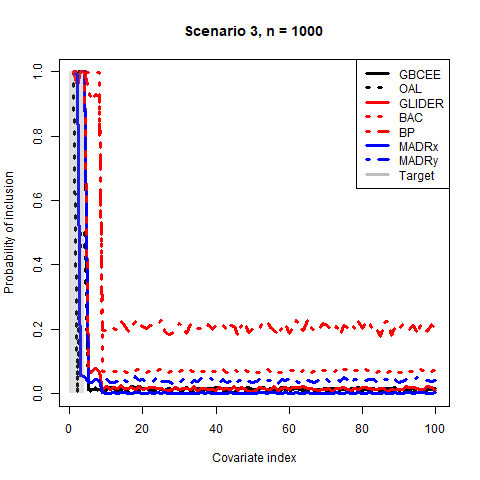}
\clearpage
\includegraphics{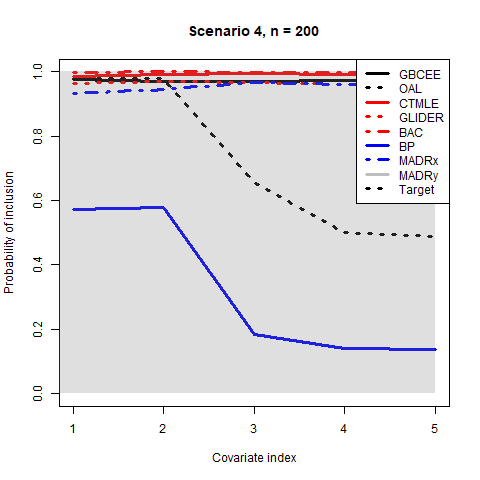}
\clearpage
\includegraphics{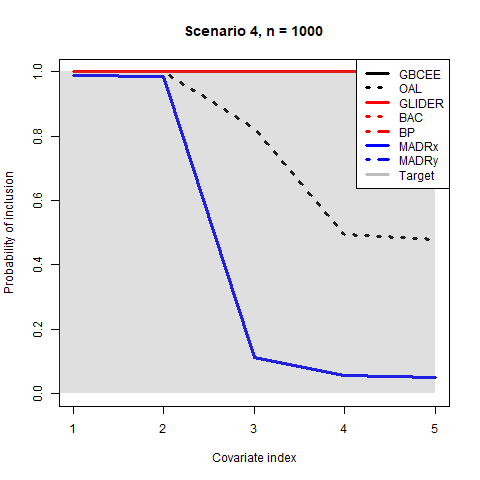}
\clearpage
\includegraphics{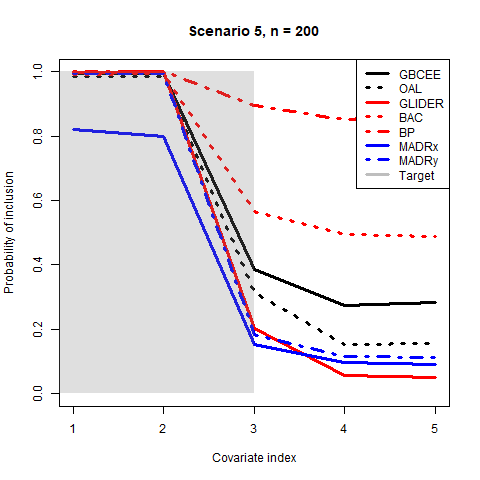}
\clearpage
\includegraphics{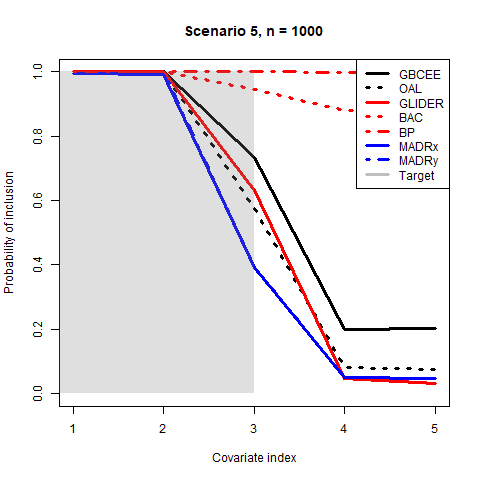}

\clearpage

\section*{Web Appendix D - Additional information on the application}

\begin{table}[!htbp] \centering 
  \caption{Descriptive characteristics of the participants in the Study of Osteoporotic Fractures according to exposure to physical activity recommandations. Numbers are mean (standard deviation) unless otherwise stated. } 
   
\begin{adjustbox}{totalheight=\textheight-6\baselineskip}
\begin{tabular}{@{\extracolsep{5pt}} lccc} 
\hline \\[-1.8ex] 
 & Unexposed & Exposed & Standardized mean difference \\ 
\hline \\[-1.8ex] 
n &   2984 &    474 &  \\ 
Body mass index (in kg/m$^2$) &  26.45 (4.37) &  25.15 (4.04) &  0.309 \\ 
Waist to hip ratio &   0.81 (0.06) &   0.80 (0.06) &  0.183 \\ 
Short Mini Mental Status Exam &  24.68 (1.61) &  24.95 (1.33) &  0.184 \\ 
Clinic &    &     &  0.267 \\ 
\ \ \ \ 1 &    751 (25.2)  &    153 (32.3)  &  \\ 
\ \ \ \ 2 &    595 (19.9)  &    122 (25.7)  &  \\ 
\ \ \ \ 3 &    780 (26.1)  &    103 (21.7)  &  \\ 
\ \ \ \ 4 &    858 (28.8)  &     96 (20.3)  &  \\ 
Age &  71.67 (5.18) &  70.07 (4.31) &  0.335 \\ 
Highest grade of school completed &  12.56 (2.70) &  13.62 (2.76) &  0.388 \\ 
Northern European -- n (\%) &   1845 (61.8)  &    298 (62.9)  &  0.021 \\ 
Central European -- n (\%) &   1512 (50.7)  &    235 (49.6)  &  0.022 \\ 
Southern European -- n (\%) &    198 ( 6.6)  &     29 ( 6.1)  &  0.021 \\ 
Jewish -- n (\%) &     56 ( 1.9)  &     17 ( 3.6)  &  0.105 \\ 
Native American -- n (\%) &     21 ( 0.7)  &      1 ( 0.2)  &  0.073 \\ 
Russian -- n (\%) &     83 ( 2.8)  &     11 ( 2.3)  &  0.029 \\ 
Other origin -- n (\%) &    121 ( 4.1)  &     16 ( 3.4)  &  0.036 \\ 
Hysterectomy -- n (\%) &    759 (25.4)  &    117 (24.7)  &  0.017 \\ 
Ovary removed -- n (\%) &    756 (25.3)  &    123 (25.9)  &  0.014 \\ 
Fracture before 50 -- n (\%) &    997 (33.4)  &    145 (30.6)  &  0.060 \\ 
Fell in past year -- n (\%) &    798 (26.7)  &    128 (27.0)  &  0.006 \\ 
Injury from fall in past year -- n (\%) &    562 (18.8)  &     80 (16.9)  &  0.051 \\ 
Fear of falling -- n (\%) &   1323 (44.3)  &    131 (27.6)  &  0.353 \\ 
Time/year of moderate physical activity at 50 &  16.58 (45.32) &  50.24 (77.91) &  0.528 \\ 
Time/year of high physical activity at 50 &   1.93 (16.05) &   7.89 (29.92) &  0.248 \\ 
Time/year of moderate physical activity at 30 &  16.81 (46.74) &  38.70 (74.39) &  0.352 \\ 
Time/year of high physical activity at 30 &   3.28 (19.89) &   7.86 (30.47) &  0.178 \\ 
Time/year of moderate physical activity when teenager &  46.61 (80.66) &  69.87 (94.15) &  0.265 \\ 
Time/year of high physical activity when teenager &  16.20 (48.09) &  20.68 (51.64) &  0.090 \\ 
Bed ridden for 7 days in past year -- n (\%)  &    126 ( 4.2)  &     11 ( 2.3)  &  0.107 \\ 
Smoking -- n (\%) &    &     &  0.080 \\ 
\ \ \ \ Never&   1887 (63.2)  &    291 (61.4)  &  \\ 
\ \ \ \ Past &    803 (26.9)  &    143 (30.2)  &  \\ 
\ \ \ \ Current &    294 ( 9.9)  &     40 ( 8.4)  &  \\ 
Caffeine intake (mg/day) & 134.06 (137.32) & 138.69 (136.24) &  0.034 \\ 
Any alcoholic beverage in past year -- n (\%) &   2046 (68.6)  &    376 (79.3)  &  0.247 \\ 
At least 1 alcoholic beverage in past 30 days -- n (\%) &   1576 (52.8)  &    314 (66.2)  &  0.276 \\ 
How often $>$3 drinks/day in past 30 days &   0.18 (0.76) &   0.22 (0.73) &  0.050 \\ 
Osteoporosis -- n (\%) &    426 (14.3)  &     39 ( 8.2)  &  0.192 \\ 
Diabetes -- n (\%) &    174 ( 5.8)  &     19 ( 4.0)  &  0.084 \\ 
Ever had a stroke -- n (\%) &     84 ( 2.8)  &      4 ( 0.8)  &  0.147 \\ 
Ever had hypertension -- n (\%) &   1114 (37.3)  &    142 (30.0)  &  0.157 \\ 
Parkinson's disease -- n (\%) &     12 ( 0.4)  &      1 ( 0.2)  &  0.035 \\ 
Arthritis -- n (\%) &   1772 (59.4)  &    253 (53.4)  &  0.121 \\ 
Stayed in hospital overnight in past year -- n (\%) &    311 (10.4)  &     39 ( 8.2)  &  0.076 \\ 
Pain around hip for most days in a month in past year -- n (\%) &   1015 (34.0)  &    146 (30.8)  &  0.069 \\ 
Thiazide use -- n (\%) &    &     &  0.116 \\ 
\ \ \ \ Never &   1971 (66.1)  &    335 (70.7)  &  \\ 
\ \ \ \ Past &    244 ( 8.2)  &     40 ( 8.4)  &  \\ 
\ \ \ \ Current &    769 (25.8)  &     99 (20.9)  &  \\ 
Non-thiazide diuretice use -- n (\%) &    &     &  0.167 \\ 
\ \ \ \ Never &   2750 (92.2)  &    455 (96.0)  &  \\ 
\ \ \ \ Past &     79 ( 2.6)  &      8 ( 1.7)  &  \\ 
\ \ \ \ Current &    155 ( 5.2)  &     11 ( 2.3)  &  \\ 
Benzodiazapine use in past year -- n (\%) &    414 (13.9)  &     63 (13.3)  &  0.017 \\ 
Sedative hypnotic use in past year -- n (\%) &     37 ( 1.2)  &     13 ( 2.7)  &  0.108 \\ 
Antidepressants use in past year -- n (\%) &    101 ( 3.4)  &     19 ( 4.0)  &  0.033 \\ 
Oral estrogen use -- n (\%) &    &     &  0.286 \\ 
\ \ \ \ Never &   1913 (64.1)  &    238 (50.2)  &  \\ 
\ \ \ \ Past &    743 (24.9)  &    158 (33.3)  &  \\ 
\ \ \ \ Current &    328 (11.0)  &     78 (16.5)  &  \\ 
Progestin use -- n (\%) &    &     &  0.171 \\ 
\ \ \ \ Never &   2819 (94.5)  &    429 (90.5)  &  \\ 
\ \ \ \ Past &     76 ( 2.5)  &     14 ( 3.0)  &  \\ 
\ \ \ \ Current &     89 ( 3.0)  &     31 ( 6.5)  &  \\ 
Difficulty walking 2 or 3 blocks -- n (\%) &    356 (11.9)  &     18 ( 3.8)  &  0.306 \\ 
Back pain in past year -- n (\%) &   1922 (64.4)  &    274 (57.8)  &  0.136 \\ 
Use walking aids -- n (\%) &    114 ( 3.8)  &     11 ( 2.3)  &  0.087 \\ 
Problems that prevent getting up or walking up stairs -- n (\%) &    283 ( 9.5)  &     27 ( 5.7)  &  0.143 \\ 
Comparative self-ratted health (1 = Excellent, 5 = Very poor)  &   1.84 (0.70) &   1.53 (0.63) &  0.461 \\ 
Not married -- n (\%) &   1500 (50.3)  &    182 (38.4)  &  0.241 \\ 
Years since menopause &  23.76 (8.03) &  21.04 (7.29) &  0.354 \\ 
Mother ever had a fracture -- n (\%) &    974 (32.6)  &    158 (33.3)  &  0.015 \\ 
Father ever had a fracture  -- n (\%) &    626 (21.0)  &    106 (22.4)  &  0.034 \\ 
\hline \\[-1.8ex] 
\end{tabular}
\end{adjustbox} 
\end{table}


\vspace*{-8pt}


\clearpage

\begin{thebibliography}{}

\bibitem{Belloni} Belloni, A., Chernozhukov, V., and Hansen, C. (2014) Inference on treatment effects after selection among high-dimensional controls. \textit{Review of Economic Studies} \textbf{81(2),} 608--650.

\bibitem{Bonaiuti} Bonaiuti, D., Shea, B., Iovine, R., Negrini, S., Kemper, H. H. C. G. et al. (2002). Exercise for preventing and treating osteoporosis in postmenopausal women. \textit{Cochrane Database of Systematic Reviews} \textbf{2,} Art. No.: CD000333. 

\bibitem{Brookhart} Brookhart, M. A., Schneeweiss, S., Rothman, K. J., Glynn, R. J., Avorn, J., and St\"urmer, T. (2006). Variable selection for propensity score models. \textit{American Journal of Epidemiology} \textbf{163,} 1149--1156.

\bibitem{Cefalu} Cefalu, M., Dominici, F., Arvold, N., and Parmigiani, G. (2017) Model averaged double robust estimation.
\textit{Biometrics} \textbf{73(2),} 410--421.

\bibitem{Dawid} Dawid, A. P. (1970) On the limiting normality of posterior distributions. \textit{Proceedings of the Cambridge Philosophical Society} \textbf{67(6),} 25--33.

\bibitem{deLuna} de Luna, X., Waernbaum, I., and Richardson, T. S. (2011). Covariate selection for the nonparametric estimation of an average
treatment effect. \textit{Biometrika} \textbf{ 98,} 861--875.

\bibitem{Diaz} D\'iaz, I., van der Laan, M. (2012). Population intervention causal effects based on stochastic interventions. \textit{Biometrics} \textbf{ 68,} 541--549.

\bibitem{Efron} Efron, B. (2013) Estimation and accuracy after model selection. \textit{Journal of the American statistical association} \textbf{109(507),} 991--1007. 

\bibitem{Ertefaie} Ertefaie, A., Asgharian, M., and Stephens, D. A. (2018) Variable Selection in Causal Inference using a Simultaneous Penalization Method. \textit{Journal of Causal Inference} \textbf{6(1),} 1--16. 

\bibitem{Haughton} Haughton, D. M. A (1988). On the choice of a model to fit data from an exponential family. \textit{The Annals of Statistics} \textbf{14(1),} 342--365.

\bibitem{Koch} Koch, B., Vock, D. M., and Wolfson, J. (2018) Covariate selection with group lasso and double robust estimation of causal effects. \textit{Biometrics} \textbf{74(1),} 8--17.

\bibitem{Looker} Looker, A. C., and Frenk, S. M. (2015) Percentage of adults aged 65 and over with osteoporosis or low bone mass at the femur neck or lumbar spine: United States, 2005-2010. \textit{National Center for Health Statistics}. URL:https://www.cdc.gov/nchs/data/hestat/osteoporsis/osteoporosis2005\_2010.htm retrieved March 24 2020.

\bibitem{Madigan} Madigan, D., York, J., Allard, D. (1995) Bayesian graphical models for discrete data. \textit{International Statistical Review} \textbf{63(2),} 215--32.

\bibitem{Pearl2009} Pearl, J. (2009) \textit{Causality: models, reasoning, and inference, 2nd ed.} New York: Cambridge University Press.

\bibitem{Pearl2011} Pearl, J. (2011) Invited commentary: Understanding bias amplification. \textit{Amercian journal of epidemiology} \textbf{174(11),} 1223--1227.

\bibitem{Porter} Porter, K. E., Gruber, S., van der Laan, M. J., Sekhon, J.S. (2011) The relative performance of targeted maximum likelihood estimators. \textit{International Journal of Biostatistics} \textbf{7(1),} 1--34.

\bibitem{Raftery} Raftery, A. E. Bayesian model selection in structural equation models. In \textit{Testing Structural Equation Models} (K. Bollen and J. Long, eds.) 163--180. Sage, Newbury Park, CA. 

\bibitem{Raftery2018} Raftery, A. E., Hoeting, J., Volinsky, C., Painter, I., Yeung, K. Y. (2018) BMA: Bayesian Model Averaging. \textit{R package version 3.18.9} https://CRAN.R-project.org/package=BMA

\bibitem{Rosenblum} Rosenblum, M., van der Laan, M. J. (2010) Targeted maximum likelihood estimation of the parameter of a marginal structural model. \textit{Biostatistics} \textbf{6(2),} Article 19. 

\bibitem{Scharfstein} Scharfstein, D. O., Rotnitzky, A., Robins, J. M. (1999) Adjusting for non-ignorable drop-out using semi-parametric nonresponse models (with discussion and rejoinder). \textit{Journal of the American Statistical Association} \textbf{94(448),} 1096-1120 (1121-1146).

\bibitem{Shortreed} Shortreed, S. M., and Ertefaie, A. (2017) Outcome-adaptive lasso: Variable selection for causal inference. \textit{Biometrics} \textbf{73(4),} 1111--1122.

\bibitem{Siddique} Siddique, A. A., Schnitzer, M. E., Bahamyirou, A., Wang, G., Holz, T. H., Migliori, G., et al. (2018) Causal inference with multiple concurrent medications: A comparison of methods and an application in multidrug-resistant tuberculosis. \textit{Statistical Methods in Medical Research}.

\bibitem{Sjolander} Sj\"olander, A., Vansteelandt, S. (2017) double robust estimation of attributable fractions in survival analysis. \textit{Statistical Methods in Medical Research} \textbf{26(2),} 948--969.

\bibitem{Talbot2015} Talbot, D., Lefebvre, G., and Atherton, J. (2015) The Bayesian causal effect estimation algorithm. \textit{Journal of Causal Inference} \textbf{30(2),} 207--236.

\bibitem{Talbot2019} Talbot, D., and Massamba V.K. (2019) A Descriptive Review of Variable Selection Methods in Four Epidemiologic Journals: There is Still Room for Improvement. \textit{European Journal of Epidemiology} \textbf{34(8),} 725--730. 

\bibitem{Thibaud} Thibaud, M., Bloch, F., Tournoux-Facon, C., Brèque, C., Rigaud, S., Dugué, B., and Kemoun, G. (2011) \textit{Impact of physical activity and sedentary behaviour on fall risks in older people: a systematic review and meta-analysis of observational studies} \textbf{9,} 5--15. 

\bibitem{Tsiatis} Tsiatis, A. (2007) \textit{Semiparametric theory and missing data} New-York: Springer Science \& Business Media.

\bibitem{vanderLaan2006} {v}an {d}er Laan, M. J., Rubin, D. (2006) Targeted maximum likelihood learning. \textit{The International Journal of
Biostatistics} \textbf{2(1),} article 11. 

\bibitem{vanderLaan2010} {v}an {d}er Laan, M., and Gruber, S. (2010) Collaborative double robust targeted maximum likelihood estimation. \textit{International Journal of Biostatistics} \textbf{6(17),} 

\bibitem{vanderLaan2011} {v}an {d}er Laan, M., Rose, S. (2011) \textit{Targeted Learning: Causal Inference for Observational and Experimental Data.} New York: Springer Series in Statistics.

\bibitem{Walker} Walker, A. M. (1969) On the asymptotic behavior of posterior distributions. \textit{Journal of the royal statistical society: Series B} \textbf{1969(31),} 80--88.

\bibitem{Walter} Walter, S., and Tiemeier, H. (2009). Variable selection: current practice in epidemiological studies. \textit{European Journal of Epidemiology} \textbf{24(12),} 733--736.

\bibitem{Wang2012} Wang, C., Parmigiani, G., and Dominici, F. (2012) Bayesian effect estimation accounting for adjustment uncertainty. \textit{Biometrics} \textbf{68(3),} 661–671.

\bibitem {Wang2015} Wang, C., Dominici, F., Parmigiani, G., and Zigler, C. M. (2015) Accounting for uncertainty in confounder and effect modifier selection when estimating average causal effects in generalized linear models. \textit{Biometrics} \textbf{71,} 654--665.

\bibitem{Wasserman} Wasserman, L. (2000) Bayesian model selection and model averaging. \textit{Journal of Mathematical Psychology} \textbf{44,} 92--107.

\bibitem{Wilson} Wilson, A., and Reich, B. J. (2014) Confounder selection via penalized credible regions. \textit{Biometrics} \textbf{70(4),} 852--861.

\bibitem{WHO} World Health Organization (2010) Global recommandations on physical activity for health. URL:https://www.who.int/dietphysicalactivity/publications/9789241599979/en/
\end{thebibliography}
\end{document}